\documentclass[aps,prl,twocolumn,superscriptaddress,amsmath,%
               amssymb,floatfix]{revtex4}

\usepackage[T1]{fontenc}
\usepackage[latin1]{inputenc}
\usepackage[english]{babel}

\usepackage{textcomp}
\usepackage{amsmath}
\usepackage{amsfonts}
\usepackage{amssymb}

\usepackage{graphicx}
\usepackage{xcolor}

\usepackage{epstopdf}

\usepackage{units}

\usepackage{ifpdf}
\ifpdf
  \newcommand*{\hyperrefcolor}{blue}
  \usepackage[%
    pdfsubject={Preprint},%
    pdftitle={Thorium-229 Search},%
    pdfauthor={misc.},%
    bookmarksopen=true,   
    plainpages=false,     
    colorlinks,           
    linkcolor=\hyperrefcolor,%
    anchorcolor=\hyperrefcolor,%
    citecolor=\hyperrefcolor,%
    urlcolor=\hyperrefcolor,%
    menucolor=\hyperrefcolor,%
    filecolor=\hyperrefcolor%
  ]{hyperref}
\fi

\usepackage[textfontcmd=footnotesize]{combinedgraphics}

\newcommand*\thor{\ensuremath{^{229}\text{Th}}}
\newcommand*\uran{\ensuremath{^{233}\text{U}}}
\newcommand*\lisaf{\ensuremath{\text{LiSrAlF}_6}}
\newcommand*\ecube{\times \unit[10^6]{eV^3}\unit{s} / (\hbar\omega)^3}

\begin{document}
\renewcommand*\includecombinedgraphics[2][]{\includegraphics{#2}}

\title{Results of a direct search using synchrotron radiation for the low-energy
       \thor{} nuclear isomeric transition}

\author{Justin Jeet}
\thanks{These authors contributed equally to this work.}
\author{Christian Schneider}
\thanks{These authors contributed equally to this work.}
\author{Scott T. Sullivan}
\thanks{Present address: AOSense, Inc., Sunnyvale, CA 94085}
\author{Wade G. Rellergert}
\thanks{Present address:  Jet Propulsion Laboratory, Pasadena, California, 91109}
\affiliation{Department of Physics and Astronomy, University of California, Los Angeles, California 90095, USA}
\author{Saed Mirzadeh}
\affiliation{Nuclear Security and Isotope Technology Division, Oak Ridge National Laboratory, Oak Ridge, Tennessee 37831, USA}
\author{A. Cassanho}
\author{H. P. Jenssen}
\affiliation{AC Materials, Inc., 756 Anclote Road, Tarpon Springs, Florida 34689, USA}
\author{Eugene V. Tkalya}
\affiliation{Skobeltsyn Institute of Nuclear Physics, Lomonosov Moscow State University, Leninskie gory, Moscow 119991, Russia}
\affiliation{Nuclear Safety Institute of Russian Academy of Science, Bol'shaya Tulskaya 52, Moscow 115191, Russia}
\author{Eric R. Hudson} 
\affiliation{Department of Physics and Astronomy, University of California, Los Angeles, California 90095, USA}

\email{christian.schneider@physics.ucla.edu}
\date{\today}

\begin{abstract}
We report the results of a direct search for the \thor{}
($I^{p} = 3/2^+\leftarrow 5/2^+$) nuclear isomeric transition, performed by exposing
\thor{}-doped \lisaf{} crystals to tunable vacuum-ultraviolet synchrotron
radiation and observing any resulting fluorescence.
We also use existing nuclear physics data to establish a range of possible
transition strengths for the isomeric transition.
We find no evidence for the thorium nuclear transition between
$\unit[7.3]{eV}$ and $\unit[8.8]{eV}$ with transition lifetime
$\unit[(1\text{--}2)]{s} \lesssim \tau \lesssim \unit[(2000\text{--}5600)]{s}$.
This measurement excludes roughly half of the favored transition search area
and can be used to direct future searches.
\end{abstract}

\maketitle

Almost four decades ago, the existence of a low-lying nuclear excited state in
\thor{} was indirectly established through the spectroscopy of $\gamma$-rays
emitted following the $\alpha$-decay of \uran{} \cite{Kroger1976}.
Subsequent indirect measurements placed this excited, isomeric state
($I^p = 3/2^+$) to be $\unit[(3.5 \pm 1.0)]{eV}$  above the ground state
($I^p = 5/2^+$) \cite{Helmer1994}.
The prospects of a laser-accessible nuclear transition touched off a flurry of
proposals to utilize this apparently unique nuclear transition as a sensitive probe of both nuclear structure and chemical environment \cite{Tkalya1996}, to constrain the variability of the fundamental constants \cite{Flambaum2006, Flambaum2009, Litvinova2009}, to check the exponentiality of the decay law of an isolated metastable state \cite{Dykhne1998}, to realize a qubit with extraordinary features  \cite{Raeder2011}, to demonstrate a gamma-ray laser \cite{Tkalya2011}, and
to construct a clock with unprecedented performance
\cite{Peik2003, Rellergert2010a, Campbell2012}.

However, as these applications generally required probing the nuclear
transition with a narrowband laser system, it was necessary to first more
precisely determine the transition energy.
Therefore, several efforts were undertaken to spectroscopically resolve the
expected ultraviolet (UV) emission from this magnetic dipole (M1) transition,
where the excited state was typically expected to be populated in the $\alpha$-decay of
\uran{} \cite{Barci2003}.
Despite initial claims of observation \cite{Irwin1997,Richardson1998}, these
searches were unsuccessful \cite{Utter1999,Shaw1999,Moore2004}.

In 2007, using a significantly improved $\gamma$-ray spectrometer, the energies of
the $\gamma$-rays emitted following \uran{} $\alpha$-decay were
remeasured.
The \thor{} isomeric transition was found to actually be in the
vacuum ultraviolet (VUV) portion of the electromagnetic spectrum \cite{Beck2007},
with an energy of $\unit[(7.8 \pm 0.5)]{eV}$ \cite{Beck2009},
thus explaining why previous searches, which used VUV insensitive detection
methods, failed, and re-energizing the community in the search for direct
observation of this nuclear transition.
And recently, the result of a search for VUV emission from the nuclear excited
state, again expected to be populated in the $\alpha$-decay of \uran{}, suggests that
the transition energy is $< \unit[7.75]{eV}$ \cite{Zhao2012}; though it is the
subject of controversy \cite{Peik2013}.

As experiments that aim to measure the nuclear transition energy via the decay
of \uran{} cannot control or modulate the observed signal, it can be difficult
to differentiate whether the observed signal is indeed from the \thor{}
isomeric transition or from known systematic effects
\cite{Utter1999,Shaw1999,Peik2013}.
Therefore, it is desirable to perform a direct measurement, where the nuclear
transition is excited by an external source of electromagnetic radiation and
the resulting fluorescence monitored, to conclusively measure the energy of the
nuclear transition.
However, given that the nuclear transition is expected to have a narrow
linewidth and is only constrained to within a $\unit[40]{nm}$ band in the VUV
region of the electromagnetic spectrum, such a direct search is daunting. 

In 2008, we proposed an experiment to utilize a VUV transparent crystal doped
with \thor{} to provide a high density sample suitable for a search using a
broadband synchrotron light source, which can easily tune over the requisite
$\unit[40]{nm}$ portion of the VUV spectrum \cite{Rellergert2010a}.
Here, we report the first results of this direct  search.
We also use existing nuclear physics data to establish a range of possible
transition strengths for the isomeric transition.
In this search, we find no evidence for the thorium nuclear transition between
$\unit[7.3]{eV}$ and $\unit[8.8]{eV}$ with transition lifetime
$\unit[(1\text{--}2)]{s} \lesssim \tau \lesssim \unit[(2000\text{--}5600)]{s}$.
Assuming no systematic effects are responsible for this lack of observation,
this result excludes roughly half of the defined transition search area and can
be used to direct future searches.

In the remainder of this manuscript, we first outline the experimental
apparatus and protocol, present sample data, and interpret the data to place
constraints on the transition strength as a function of transition energy.
We also establish a range for the possible transition energy and lifetime, and
conclude with a discussion and estimation of possible systematic effects.

\begin{figure}
  \centering
  \includecombinedgraphics[vecwidth=\hsize,keepaspectratio]%
    {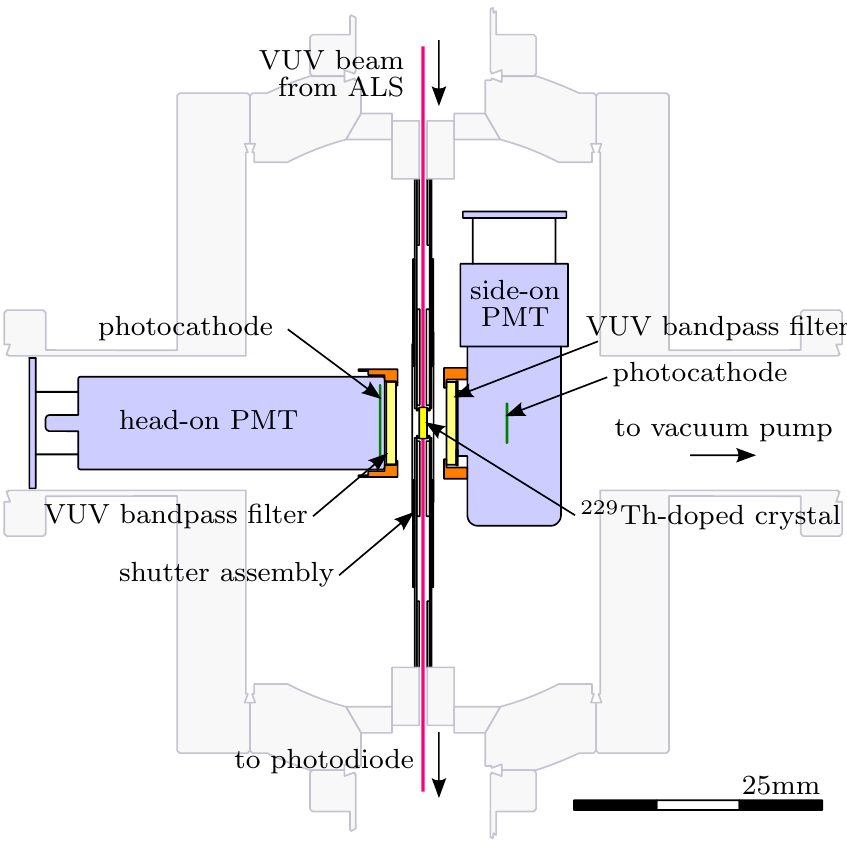}
  \caption{Schematic of the experimental setup.
    A \thor{}-doped crystal is placed in the center of a
    $\unit[4]{mm}$ thick mount. The crystal is illuminated with synchrotron radiation along its long axis.
    Any resulting fluorescence is collected perpendicular to this axis by a head-on type and side-on
    type PMT. The head-on PMT can be optionally
    cooled to $\unit[-20]{^\circ C}$.
    Low-profile mechanical shutters (UniBlitz) shield the PMTs
    from scattered light during the VUV illumination.
    The assembly is held under a
    pressure $< \unit[10^{-5}]{mbar}$.}
  \label{schematic}
\end{figure}

Conceptually, the experimental apparatus, shown in Fig.~\ref{schematic}, and
the protocol are simple.
A VUV transparent, \thor{}-doped crystal is illuminated for time $T_\text{e}$
with VUV photons.
If the VUV photons are resonant with the isomeric transition, a fraction of the
\thor{} nuclei are excited to the ($I^{p} = 3/2^+$) isomeric state.
Following the excitation period, the VUV photon source is shuttered and any
subsequent fluorescence, which results from spontaneous decay back to the
($I^{p} = 5/2^+$) ground state, is recorded by opening two low-profile
shutters to expose two photomultiplier tubes (PMTs) to the crystal for a time
$T_\text{d}$.
This simple approach is complicated by the experimental realities of generating
tunable VUV light, the availability of high-purity \thor{}, and the construction of a \thor{}-doped, VUV transparent
crystal. 

\begin{figure}
  \centering
  \includecombinedgraphics[vecwidth=\hsize,keepaspectratio]{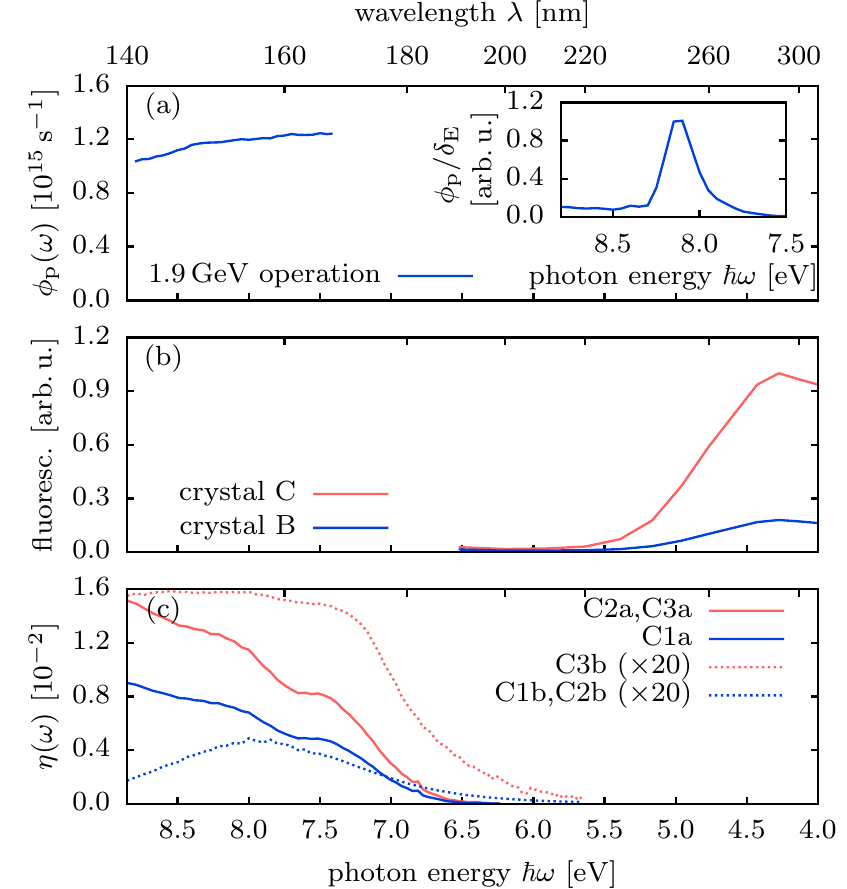}
  \caption{Photon energy (wavelength) dependence of relevant experimental
    parameters:
    (a) VUV photon flux $\phi_\text{p}$ of the ALS beamline
    9.0.2.1 for normal ($\unit[1.9]{GeV}$) operation; inset shows sample ALS spectral lineshape taken with monochromator resolution $\delta_\text{E} \approx \unit[0.05]{eV}$ (courtesy of Oleg Kostko).
    (b) UV fluorescence spectrum of the crystals due to
    radioactive decay, recorded with a McPherson 234/302
    monochromator and a Hamamatsu R1527P PMT.
    (c) Total detection efficiencies $\eta$ of different
    configurations of the detection system (see Tab.~\ref{config}).}
  \label{parameters}
\end{figure}

For a tunable VUV light source, we utilize beamline 9.0.2.1 at the
Advanced Light Source (ALS) synchrotron \cite{Heimann1997}.
In normal operation ($\unit[1.9]{GeV}$ electron energy), this beamline can be
tuned within $\hbar \omega \approx \unit[(7.4\text{--}30)]{eV}$.
We characterize the photon flux  $\phi_\text{p}$ of the beamline with a
Opto Diode Corporation (ODC) SXUV-100 VUV damage resistant photodiode,
calibrated against a NIST-calibrated ODC AXUV-100G photodiode, and find
$\phi_\text{p} = \unit[(1.0\text{--}1.25) \times 10^{15}]{s^{-1}}$
(see Fig.~\ref{parameters}(a)).
The VUV spectrum is nearly Lorentzian with linewidth
$\hbar \Gamma \approx \unit[0.19]{eV}$ and exhibits a constant
``tail'' up to $\approx \unit[10]{eV}$ (see inset Fig.~\ref{parameters}(a)).
We reduce the photon flux to $\xi \approx 0.7$ of its measured value to account
for the fraction of ALS photons which are contained in this tail.

The \thor{} dopant used in this work was purchased from Isotope Distribution Office, U.S. Department of Energy and was extracted from  ORNL mass-separated $^{233}$U, batch UTHX001, which contains only ppb levels of $^{232}$U.  The mass distribution of thorium isotopes in this sample was:  \thor{}~(75.6\%), $^{228}$Th (<~0.1\%), $^{230}$Th (0.48\%) and $^{232}$Th (23.8\%).  

The development of a suitable \thor{}-doped crystal has been one of the most
significant challenges of this work and is described elsewhere
\cite{Rellergert2010b, Hehlen2013}.
Briefly, over the course of the last six years, we have grown and tested a
variety of VUV transparent crystals doped with the common $^{232}$Th isotope to
find a material which satisfies all of the requirements of this experiment,
which include: high VUV transparency; absence of long-lived fluorescence,
resulting from e.g.{} color center formation; resistance to the effects of
nuclear radiation; and low radiation-induced scintillation. 
From this work, we settled on \lisaf{} and have produced three
\thor{}-doped \lisaf{} crystals with dimensions $\approx \unit[3]{mm} \times
\unit[3]{mm} \times \unit[10]{mm}$.
The amount of \thor{} in each crystal was assayed by $\gamma$-ray
spectroscopy (Ortec GMX-50220-P) \cite{Alexiev2002} and found to be
$\unit[(115\pm5)]{nCi}$, $\unit[(290\pm40)]{nCi}$, and $\unit[(2.2\pm3)]{\mu Ci}$.
In this work, we use only the latter two crystals, referred to as crystal B and
C, which have a thorium atomic density of
$n_\text{Th} \approx \unit[5.8 \times 10^{16}]{cm^{-3}}$ and
$n_\text{Th} \approx \unit[4.1 \times 10^{17}]{cm^{-3}}$, respectively.

Due to scintillation following the radioactive decay of \thor{}
($\unit[4.8]{MeV}$ $\alpha$-decay with a half-life of $\unit[(7917\pm48)]{y}$~\cite{Varga2014}) and
its daughter isotopes, the crystals emit photons in the UV region.
The spectrum of the scintillation in each crystal is shown in
Fig.~\ref{parameters}(b).
All spectra show a maximum at $\approx \unit[300]{nm}$ and drop to
$\unit[50]{\%}$ at $\sim\unit[250]{nm}$.
This UV fluorescence together with scintillation induced directly in the PMTs
by the crystal radioactivity is the dominant background in our measurements.

At a given ALS beam energy, $\hbar\omega$, in the limit of weak excitation, the number of
detected photons from the nuclear fluorescence, $N_d$, in a time interval $\left\{t_1,t_2\right\}$ is given as: 
\begin{equation}
  \begin{gathered}
    N_d = N_0
      \left(1 - \text{e}^{-T_\text{e} / \tau'}\right)
      \left(\text{e}^{-t_1 / \tau'}
      - \text{e}^{-(t_1 + t_2)/ \tau'}\right) \\
    \text{with} \quad N_0 = \frac{2}{3} \, \eta(\omega) \,
      \frac{\lambda_0^2}{2\pi}
      \, \frac{\xi\phi_\text{p}'(\omega) \, n_\text{Th} \, l}{\Gamma_\text{B} + \Gamma}
      \frac{1}{1+4\left(\frac{\omega-\omega_0}
      {\Gamma_\text{B} + \Gamma}\right)^2},
  \label{nphot}
  \end{gathered}
\end{equation}
where $\tau'$ is the lifetime of the \thor{} transition inside the crystal, $\phi_\text{p}'$ is the photon flux transmitted through the crystal, 
$\eta(\omega)$ is the
total efficiency of the detection system, $\omega_0$ ($\lambda_0$) is the
unknown isomeric transition energy (wavelength), $l$ is the length of the
crystal, and $\Gamma_\text{B} \approx \unit[10]{kHz}$  is the inhomogeneously
broadened linewidth in the crystal environment \cite{Rellergert2010b}.

\begin{table}
  \centering
  \newbox\ttt\newskip\tttskip
  \def\putlow#1{\setbox\ttt=\hbox{\raise -\tttskipe\hbox{#1}}\dp\ttt=0pt\box\ttt}
  \tttskip=0.5\baselineskip
  \edef\tttskipe{\the\tttskip}
  \begin{tabular}{r@{\,}c@{\quad}c@{\quad}c@{\quad}c@{\quad}c@{\quad}c}
    \hline
    con\rlap{fig.}&~~& crystal & PMTs & filter & $d$ [$\unit{mm}$] & $\hbar\omega$ [$\unit{eV}$]\\
    \hline\hline
    \putlow{C1}&a & \putlow{C} & R6835 & no  & $\phantom{4}12$ & \putlow{$7.4\text{--}8.8\phantom{5}$}\\
               &b &            & R8486 & yes & $\phantom{4}25$ & \\
    \hline
    \putlow{C2}&a & \putlow{B} & R6835 & no  & $5\text{--}6$ & \putlow{$7.4\text{--}8.8\phantom{5}$}\\
               &b &            & R8486 & yes & $\phantom{4}25$ & \\
    \hline
    \putlow{C3}&a & \putlow{B} & R6835 & no  & $6\text{--}7$ & \putlow{$7.4\text{--}8.25$}\\
               &b &            & R7639 & no  & $\phantom{4}25$ & \\
    \hline
  \end{tabular}
  \caption{Relevant configurations of the experimental system
    (see Fig.~\ref{schematic}) and searched VUV photon energy range.
    All PMTs are manufactured by Hamamatsu, the VUV bandpass filter
    ($\unit[(150 \pm 27)]{nm}$) is an Acton Research Corporation 150-N-MF-1D,
    and $d$ denotes the distance of
    the photocathode to the crystal center.
    }
  \label{config}
\end{table}

The determination of $\eta(\omega)$ for each configuration requires
knowledge of the PMT quantum efficiency $\eta_\text{PMT}$, its solid angle
fraction $\eta_\text{SAF}$,
and the transmission of a VUV bandpass filter $\eta_\text{filter}$, optionally used to reduce
the UV background.
Both $\eta_\text{PMT}$ and $\eta_\text{filter}$ are characterized using a
McPherson 234/302 monochromator,
deuterium lamp, and the above-mentioned photodiodes and agree with the
calibrations provided by the respective manufacturers.
The different $\eta_\text{SAF}$ are
determined using commercial (Zemax) ray-tracing software and independently verified with homemade ray-tracing software.

To better discriminate any \thor{} nuclear fluorescence from spurious PMT signals,
we simultaneously employ two PMTs of differing technology for each measurement.
Specifically, three overall configurations, listed in Tab.~\ref{config}, are used in our 
measurements at the ALS and their resulting total detection efficiencies are shown in
Fig.~\ref{parameters}(c).

Finally, the PMT electronic signals are (optionally) amplified (Stanford Research SR445A)
and recorded using a digitzer (CAEN DT5751), which stores the waveform of every
detected event for off-line analysis.

Using the experimental system from Fig.~\ref{schematic}, we were given a total
of 96 hours of ALS beam time
in 8- and 16-hour shifts from August 20 to September 5, 2014.
We performed a search in the range
$\unit[(7.412\text{--}8.8)]{eV}$ with step size $\le \unit[0.1]{eV}$. 
We chose an illumination time $T_\text{e} = \unit[2000]{s}$ for each VUV
energy and a detection time $T_\text{d} = \unit[1000]{s}$
($T_\text{d} = \unit[1800]{s}$) for configurations C1 and C2 (C3).

\begin{figure}
  \centering
  \includecombinedgraphics[vecwidth=\hsize,keepaspectratio]{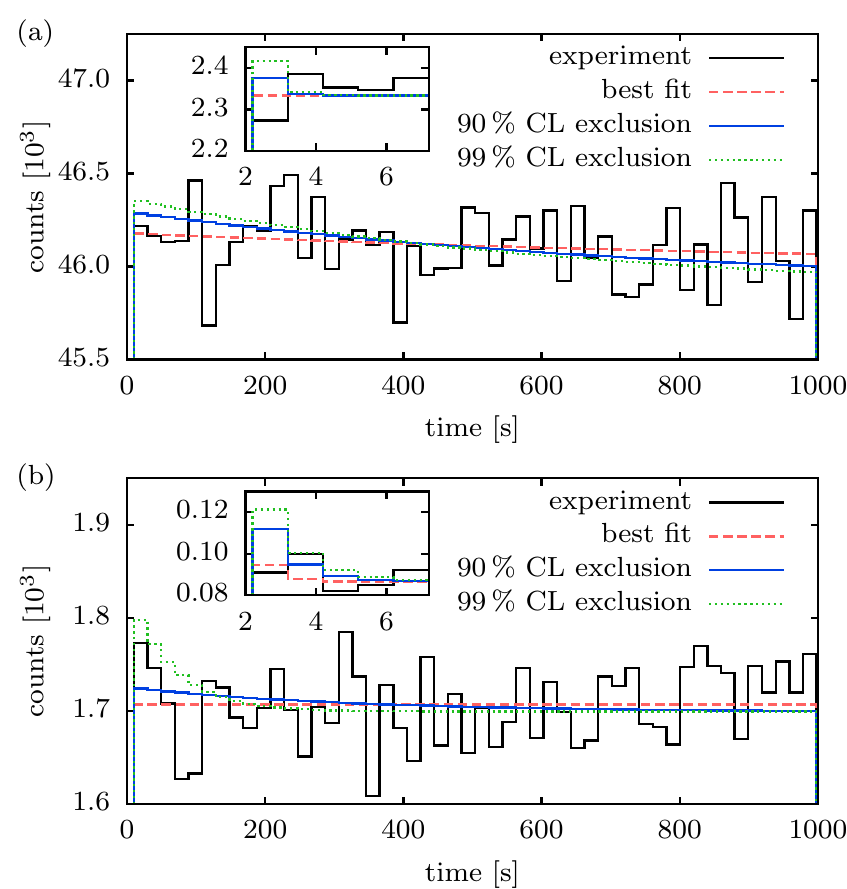}
  \caption{Representative data at a VUV energy
    $\hbar \omega = \unit[7.7]{eV}$ for (a) configuration
    C2a and (b) C2b, respectively.
    The illumination time is $T_\text{e} = \unit[2000]{s}$.
    Histograms represent the binned photon counts recorded in the time interval
    $[\unit[10]{s}; \unit[998]{s}]$ after shuttering the VUV beam
    (solid, black).
    Also shown are the best-fit result (dashed, red), the
    curve corresponding to the $\unit[90]{\%}$ CL (solid, blue), and $\unit[99]{\%}$ CL (dotted, green),
    according to Eq.~\eqref{nphot}.
    The lifetimes from the $\unit[90]{\%}$ CL cases
    are used to set an upper bound on the lifetime $\tau'$.
    Insets show the binned photon counts and fit curves in the time interval
    $[\unit[2.2]{s}; \unit[7.2]{s}]$ used to obtain the lower bound on $\tau'$.}
  \label{data}
\end{figure}

Fig.~\ref{data} shows the recorded (binned) crystal fluorescence after ALS
illumination at $\hbar \omega  = \unit[7.7]{eV}$ using configuration C2.
This data is representative of all recorded data.
From data such as these, we establish an upper bound on the transition lifetime
using Eq.~\eqref{nphot} evaluated for each bin and the Feldman-Cousins prescription \cite{Feldman1998} since our signal cannot be negative. For this analysis, we use a $\Delta\chi^2$ 
test statistic for the lifetime $\tau'$ that is profiled in the background count-rate
nuisance parameter \cite{Murphy2000}.
The first $\unit[10]{s}$ of each data trace are omitted to mitigate the
effect of any scattered light and/or short-lived UV/VIS fluorescence.
Contrarily, for the lower bound on the lifetime, we derive the average background signal
from data taken for $t > \unit[200]{s}$ after ending the VUV illumination, for
which no signal photons are expected.
Then, we perform a Feldman-Cousins analysis without a nuisance parameter using
only the first $\unit[5]{s}$ of recorded data (starting
at $\unit[2.2]{s}$ after ending the VUV illumination).
For each experimentally probed VUV energy,
we analyze the data assuming detunings of the ALS beam from the thorium 
transition by $\unit[-0.2]{eV}$ to $\unit[+0.2]{eV}$ in $\unit[0.01]{eV}$ steps and present the most 
excluding lifetimes, long and short, obtained from all configurations.

\begin{figure}
  \centering
  \includecombinedgraphics[vecwidth=\hsize,keepaspectratio]%
  {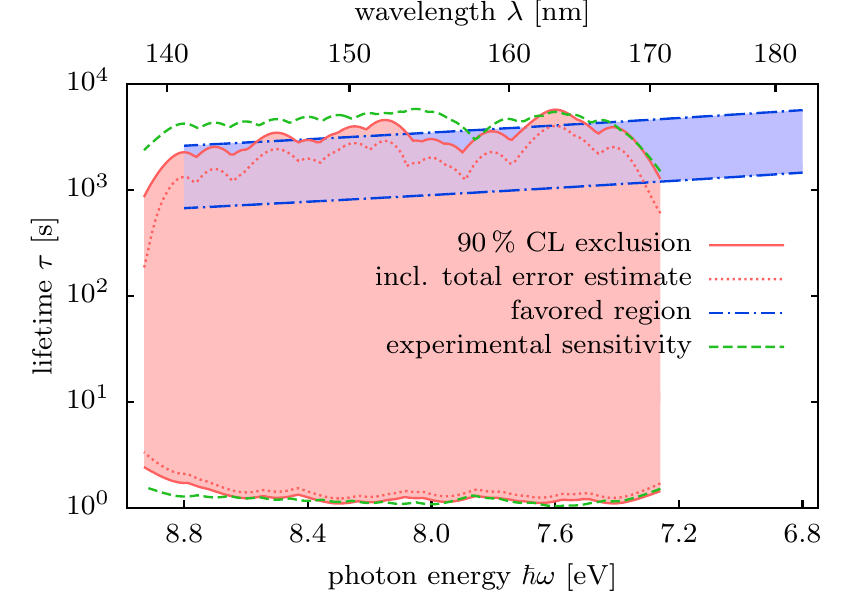}
  \caption{Exclusion region for the vacuum lifetime $\tau$ of the \thor{}
    isomeric transition as a function of photon energy $\hbar\omega$ of the
    transition.
    The red shaded area between the solid red lines can be excluded by the
    experimental data at a CL of $\unit[90]{\%}$.
    A reduction of our sensitivity by our total error budget (Tab.~\ref{errors})
    would reduce the excluded region to the area between the red dotted lines.
    For comparison, the experimental sensitivity (see text, green, dashed) and the favored
    region determined by theoretical considerations are given (blue shaded area
    between dash-dotted lines; see text).}
  \label{exclusion}
\end{figure}

Finally, since this data is recorded in a dielectric medium with refractive
index $n$, the M1 transition rate is enhanced by a factor of $n^3(\omega)$
relative to the rate in vacuum \cite{Rikken1995}---for reference, $1.46 \le n \le 1.51$ over the scanned photon energy range.
Therefore, to compare to the expected lifetime range, we convert the bounds to
vacuum lifetimes, $\tau = n^3\,\tau'$ (compare Eq.~\eqref{nphot}).
The resulting $\unit[90]{\%}$ confidence level (CL) excluded region, which comes
from configurations C2a and C3a, is shown in Fig.~\ref{exclusion}
(red shaded area between solid red lines).
The variations of the exclusion region borders (solid red lines) results from varying VUV photon fluxes, the
discrete ALS VUV energies $\hbar\omega$ used for the scan, UV background fluctuations, and
statistical fluctuations.

\begin{table}
  \centering
  \begin{tabular}{l@{\quad}c@{\quad}r}
    \hline
      parameter & symbol & rel. error\\
    \hline\hline
      ALS VUV energy           & $\hbar\omega$     & $0.02$ \\
      ALS VUV linewidth        & $\Gamma$          & $0.10$ \\
      ALS VUV spectrum         & $\xi$             & $0.40$ \\
      photon flux through crystal & $\phi_\text{p}'$  & $0.15$ \\
      solid angle fraction     & $\eta_\text{SAF}$ & $0.15$ \\
      PMT quantum efficiencies & $\eta_\text{PMT}$ & $0.15$ \\
      (eff.) crystal length    & $l$               & $0.10$ \\
      \thor{} density          & $n_\text{Th}$     & $0.15$ \\
    \hline\hline
      total                    &                   & $\approx 0.52$ \\
    \hline
  \end{tabular}
  \caption{Statistical and systematic errors for the individual contributions
    to $N_{0}$ in Eq.~\eqref{nphot}.
    We estimate bounds for the systematic error, which are significantly larger
    than statistical errors for most parameters.
    For the statistical errors, we report $1.64$ standard errors (corresponding
    to a $\unit[90]{\%}$ CL).
    The total is estimated using standard error propagation, because the
    parameters (except for $\phi'_\text{p}$ and $\eta_\text{PMT}$) are independent.
    The configurations involving a VUV filter were inferior and did not enter
    in the final results such that $\eta_\text{filter}$ is not listed.}
  \label{errors}
\end{table}

The impact of the exclusion shown in Fig.~\ref{exclusion} can be better
understood by considering the possible transition energy and lifetimes allowed
by previous experimental data.
While there is general acceptance that the measurement of
Ref.~\citenum{Beck2009} defines the possible range of the isomeric transition
energy, there is less consensus on the possible range of the isomeric transition
strength.
Ref.~\citenum{Ruchowska2006} employs a quasi-particle plus phonon-model
calculation to predict that the transition lifetime is
$\tau = 2.23 \ecube$.
Refs.~\citenum{Helmer1994}, \citenum{Beck2007}, and \citenum{Zhao2012} use the
fact that a transition between the same two Nilsson states has been observed
in \uran{} at $\unit[312]{keV}$, and predict that transition lifetime is
$\tau \approx 8\ecube$.
And, Ref.~\citenum{Dykhne1998} calculates the transition lifetime to be
$\tau = 0.66\ecube$ by finding the matrix element of
the nuclear transition in terms of another M1 transition in the thorium 
nucleus ($9/2^+5/2\left[633\right]\rightarrow7/2^+3/2\left[631\right]$),
which was previously measured \cite{Bemis1988}. 

Of these calculations, the method of Ref.~\citenum{Ruchowska2006} is accurate
to within a factor of $\sim 4$ for the cases where experimental lifetimes are
available, while the method of Refs.~\citenum{Helmer1994}, \citenum{Beck2007},
and \citenum{Zhao2012} provides only a rough estimate since nuclei specific
effects, such as the Coriolis interaction \cite{Dykhne1998}, can modify the
lifetime.
In contrast, the technique of Ref.~\citenum{Dykhne1998} has been shown to be
accurate to within experimental error in the case of other nuclei, e.g.~\uran{}
and $^{225}\text{Ra}$, where data is available.
Therefore, we prefer to compare our experimental results to this latter method,
with the modification that since the original work of Ref.~\citenum{Dykhne1998}
three new experimental measurements of the $9/2^+5/2\left[633\right]\rightarrow
7/2^+3/2\left[631\right]$ transition lifetime have been made
\cite{Gulda2002, Barci2003,Ruchowska2006}, which differ slightly from the
original measurement \cite{Bemis1988}.
Allowing for two standard deviations away from the 
mean of these measurements and two standard deviations in the measurement of
Ref.~\citenum{Beck2009}, we have constructed a ``favored region'' (bounded by
the maximum $\tau_\text{u} = 1.79\ecube$ and minimum
$\tau_\text{l} = 0.46\ecube$ isomeric state lifetime) where
the transition should lie at approximately the $\unit[90]{\%}$ CL, shown as
the (blue) shaded region between dash-dotted lines in Fig.~\ref{exclusion}. 

Also shown in Fig.~\ref{exclusion} as a dotted line is an estimate of the
potential impact of systematic errors on the calibration of the experimental
sensitivity.
This exclusion is created by reducing  $N_{0}$ by the total error
budget given in Tab.~\ref{errors}.
To be clear, the best estimate of the exclusion is given by the solid
$\unit[90]{\%}$ CL exclusion line, which is conservatively constructed using the
measured experimental parameters.
The systematic error bound is shown to give an estimate of the possible
reduction in this limit due to factors such as uncontrolled changes in the ALS linewidth
with time and beam energy.

Finally, the experimental sensitivity, defined as in Ref.~\cite{Feldman1998}, is determined with the same 
prescription as used for the bounds of the exclusion region, but by 
analyzing experimentally recorded background data without prior VUV 
illumination (Fig.~\ref{exclusion}, dashed green curve).
Our exclusion region is almost at the level of the sensitivity, indicating little excess noise.

In conclusion, barring the presence of unknown systematic effects, such as an unexpected non-radiative relaxation channel or optical trapping \cite{Rellergert2010a}, the present result excludes the existence of the \thor{} isomeric
transition with a vacuum lifetime
$\unit[(1\text{--}2)]{s} \lesssim \tau \lesssim \unit[(2000\text{--}5600)]{s}$
for the energy range
$\hbar\omega = \unit[(7.3\text{--}8.8)]{eV}$ at the $\unit[90]{\%}$ CL.
This experiment did not probe energies below $\unit[7.4]{eV}$ because in normal
operation the ALS beamline cannot reach these energies.
Future work will concentrate on improving the limits on the transition for
$\hbar\omega = \unit[(7.4\text{--}8.1)]{eV}$ and probing
$\hbar\omega < \unit[7.4]{eV}$ for the first time.
To accomplish the former will likely require the use of a VUV laser system,
which provides a higher spectral irradiance than the ALS \cite{Heimann1997, Ng2014}, as
the
detection system is already quite efficient; while the latter will require
either operation of the ALS in a special low-energy mode or a VUV laser
system.

We thank David DeMille for useful discussions;
Bruce Rude, Kevin Wilson, Musahid Ahmed, Oleg Kostko, and Sarah Ferrell for their support
at the ALS and providing the measurement of the ALS spectral lineshape; Robert Cousins for expert advice on data analysis; Alyssa Ruiz and James Hefley for help with $\gamma$-ray spectroscopy; and Richard Greco, Markus Hehlen, and Justin Torgerson for help with crystal characterization. The ALS is supported by the U.S. DOE under Contract No. DE-AC0205CH11231. This work is supported by the DARPA QuASAR program and the ARO under W911NF-11-1-0369:P00004, NIST PMG, and DOE Office of Nuclear Physics, Isotope Program.

\bibliographystyle{apsrev4-1}
\bibliography{ThoriumSearch.bib}

\end{document}